# Efficient creation of dipolar coupled nitrogen-vacancy spin qubits in diamond

**I Jakobi[1], S A Momenzadeh[1], F Fávaro de Oliveira[1], J Michl[1], F Ziem[1], M Schreck[2], P Neumann[1], A Denisenko[1] and J Wrachtrup[1]**

[1] 3. Physikalisches Institut, Unversität Stuttgart and Institute for Integrated Quantum Science and Technology IQ$^{ST}$, Pfaffenwaldring 57, 70569 Stuttgart, Germany
[2] Experimentalphysik IV, Institut für Physik, Universität Augsburg, Universitätsstraße 1 Nord, 86159 Augsburg, Germany

Email: i.jakobi@physik.uni-stuttgart.de

**Abstract**. Coherently coupled pairs or multimers of nitrogen-vacancy defect electron spins in diamond have many promising applications especially in quantum information processing (QIP) but also in nanoscale sensing applications. Scalable registers of spin qubits are essential to the progress of QIP. Ion implantation is the only known technique able to produce defect pairs close enough to allow spin coupling via dipolar interaction. Although several competing methods have been proposed to increase the resulting resolution of ion implantation, the reliable creation of working registers is still to be demonstrated. The current limitation are residual radiation-induced defects, resulting in degraded qubit performance as trade-off for positioning accuracy. Here we present an optimized estimation of nanomask implantation parameters that are most likely to produce interacting qubits under standard conditions. We apply our findings to a well-established technique, namely masks written in electron-beam lithography, to create coupled defect pairs with a reasonable probability. Furthermore, we investigate the scaling behavior and necessary improvements to efficiently engineer interacting spin architectures.

## 1. Engineering dipolar coupled spin qubits in diamond

Engineering and scaling of coupled qubit systems or quantum registers lies at the core of quantum information processing as the expansion of quantum algorithms to useful applications depends on computational resources [1]. Solid state spin systems [2,3] have the prospect of chip-scale nanofabrication, where experience can be drawn from decades of development. Here, designed structures of coherently coupled spins, e.g. chains and arrays, are promising architectures for applications like quantum processors, simulators [4] or repeaters [5], and could find utilization in fields beyond quantum computation, for example as sensing arrays in quantum metrology [6,7]. Interestingly, in nanoscale metrology coupled spin dimers and trimers would boost the performance of such sensors by exploiting entanglement leading to useful improvements [8].

Single nitrogen-vacancy defect centers (NV) in diamond are known to be promising spin qubits and hybrid quantum registers [9]. A qubit defined on the $m_S = 0, \pm 1$ sublevels of the electron spin ($S = 1$) in the negative charge state can be controlled with static and dynamic magnetic fields, e.g. driving Rabi oscillations or inducing Larmor precessions, which gives rise to single qubit gate operations on a time scale of tens to hundreds of nanoseconds. In bulk samples long spin coherence times, typically on the order of tens to hundreds of µs and ranging up to ms at room-temperature [10], allow for many gate operations before the information stored on the qubit is lost. In addition, the spin can be initialized and readout optically, which allows to conveniently address single NV defects. The Hamiltonian describing the dynamics of a single NV spin can be found in the supplementary information.

The electron spin interacts with nuclear spins in the proximity of the defect. These can be for example paramagnetic $^{13}$C (spin 1/2) in the diamond lattice or the intrinsic $^{14}$N, or $^{15}$N (spin 1 and ½, respectively). Nuclear spins can be controlled and readout via the NV electron spin through the hyperfine interaction [11,12] which allows single qubit gate operations as well as non-local gate operations between sets of nuclear spins and the central electron spin [9,13]. In addition, coherence times of nuclear spins can reach seconds at ambient conditions [14], making them an ideal storage resource for quantum information. Hence, qubits defined on the nuclear spins can expand a single NV to a versatile hybrid quantum register reaching sizes of up to four qubits [9]. However, such registers are not scalable as the coupling strength of suitable $^{13}$C spins has to be larger than the inverse coherence lifetime [15] and the spectral resolution to address individual nuclear spins is limited by the longitudinal relaxation time of the central spin [16]. Ultimately, any scaling approach has to involve coupling multiple NV electron spin nodes between which quantum information can be shared.

There are several approaches to couple multiple NV electron spins. At cryogenic temperatures NV spins can be entangled with emitted photons, which are then processed with linear optics QIP schemes [17,18], or the electron spin can be coupled to the mechanical motion of nanoscale oscillators [19]. Here, we focus on an approach that is feasible at room temperature and does not rely on mediation but rather uses direct dipolar magnetic coupling [20-22], as illustrated in Figure 1 (a). Two electron spins A and B interact through their magnetic moments according to the interaction Hamiltonian

$$\widehat{H}_{\text{dip}} = \frac{\mu_0 \, h^2 \gamma^2}{4 \pi \, r^3} \left( \vec{S}_A \cdot \vec{S}_B - 3(\vec{S}_A \cdot \vec{r})(\vec{S}_B \cdot \vec{r}) \right), \quad (1)$$

where $\mu_0$ is the vacuum permeability, $h$ is the Planck constant, $\gamma = 28.03$ GHz/T is the gyromagnetic ratio (divided by $2\pi$) of the NV electron spin, $\vec{S}_i = R_i \cdot \vec{S}'_i = R_i \cdot (\hat{S}_{x,i}, \hat{S}_{y,i}, \hat{S}_{z,i})^T \mid_{i=A,B}$ are the dimensionless spin operators (expressed in a common frame through rotation matrices $R_i$), $\vec{r}$ is the unit vector connecting them and $r$ is the distance. In the case of two NV spins the interaction can be reduced to an $\hat{S}_{z,A} \hat{S}_{z,B}$ coupling and approximated to

$$\widehat{H}_{\text{dip}} \approx h \nu_{\text{dip}} \hat{S}_{z,A} \hat{S}_{z,B}, \quad (2)$$

with the dipolar coupling strength

$$\nu_{\text{dip}}(r, \phi_{AB}, \phi_{Ar}, \phi_{Br}) = \frac{\mu_0 \, h \, \gamma^2}{4 \pi \, r^3} (\cos \phi_{AB} - 3 \cos \phi_{Ar} \cos \phi_{Br}). \quad (3)$$

$\phi_{AB}$ is the angle between the two spin quantization axes, which correspond to the crystal axes of the respective defects, and $\phi_{Ar}, \phi_{Br}$ are the angles between each quantization axis and the connecting vector $\vec{r}$ (see supplementary information for a detailed derivation). All parameters of $\nu_{\text{dip}}$ depend on the relative position and orientation of the two spins which is thus a fixed property for each pair of NVs. With its inverse cubic scaling, dipolar coupling is a short-ranged interaction. For example, a pair of NV defects at a distance of about 33 nm and a median angular contribution couples with merely 1 kHz (see Figure 1 (d)) and even with ideal orientations of the dipoles the same coupling strength requires a distance of 47 nm.

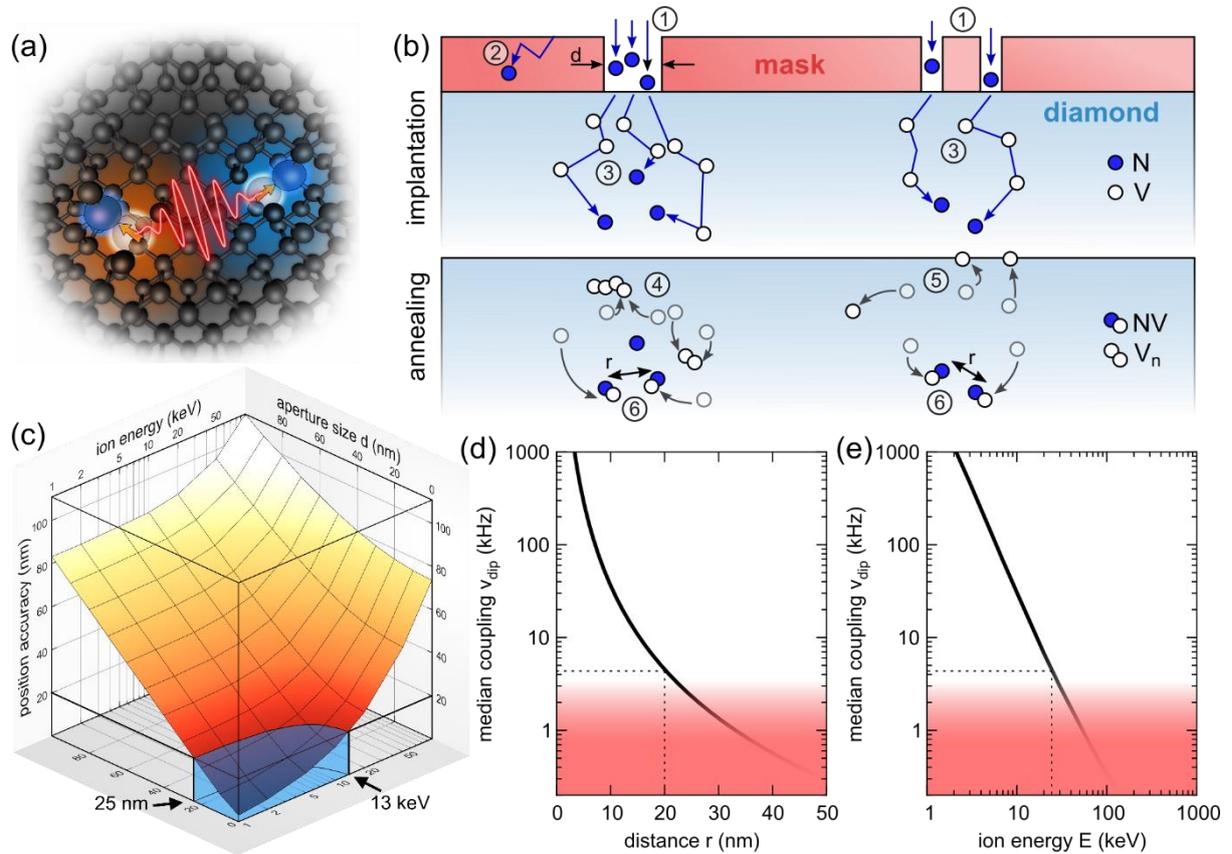

**Figure 1 (a)** Artwork of a coupling NV pair. **(b)** Schematic of nanomask implantation – Nitrogen ions (blue circles) are accelerated towards a diamond sample (1). A mask is able to block nitrogen radiation except for apertures of diameters $d$ on the nano-scale (2). Ions passing through a single (left) or targeted through individual (right) apertures penetrate the diamond surface and scatter with lattice atoms (straggling), creating vacancies (white circles) in their tracks (3). The positioning accuracy depends on the aperture size and the straggling. During annealing vacancies become mobile while nitrogen atoms are stationary. Vacancies can form agglomerates (4), recombine at the surface or in the lattice (5) or combine with nitrogen atoms to form NV defects (6). **(c)** Positioning accuracy in terms of full width half maximum of the resulting ion distribution (ion resolution) depending on ion energy and aperture size $d$. The blue cut denotes the parameter range with an accuracy below 20 nm. **(d)** Coupling strength of two NV spins over their distance with a median angular contribution. With typical coherence times of implanted NVs ranging up to several hundreds of µs, couplings below a few kHz are unlikely to result in strong coupling (red gradient). This translates to an approximate threshold of 20 nm for the strong coupling regime. **(e)** Simulated median coupling strength of two NVs, implanted through the same point shaped aperture, depending on the ion energy.

Note that throughout the work we often use median instead of mean values as many simulated and experimental statistics depend on random sampling where the median is less prone to be influenced by outliers and better represents central tendencies.

In general, any pair of spins interacts through their magnetic moments, whilst in practice the distance is limited by the coherence times as the coupling can only have an effect if it acts on a shorter timescale than the relaxation. Typical detection schemes of the coupling involve preparing one spin (sensor) in a superposition state with a phase that is sensitive to the magnetic field originating from the second spin (emitter) which is prepared in a magnetic state $|m_S = \pm 1\rangle$. The detection is then limited by the transverse relaxation time $T_2$ of the sensor, as well as the longitudinal relaxation time $T_1$ of the emitter.

In order to exchange quantum information between the spins, i.e. create entangled states between two spins, both need to act as sensors at the same time. If only one spin fulfills the requirement they can only be classically correlated which we dub "weak coupling" in the following. The limitation of a quantum register consisting of two NV nodes can thus be expressed in the "strong coupling" condition

$$\nu_{\text{dip}} > \frac{1}{\min(T_{2A}, T_{2B})}. \tag{4}$$

This is the primary limitation for the scaling of the NV quantum register via dipolar coupling. In essence, NVs need to be created at close distances to produce large couplings, while still having an unperturbed environment to retain long coherence times. Based on record coherence times of ms, the previously mentioned example of $\nu_{\text{dip}}$=1 kHz hereby marks a threshold where pairs with distances larger than 50 nm will not fulfill the condition. With typical coherence times of tens to hundreds of μs, distances as small as 10-20 nm are required. Note that the effective coupling strength is four times larger if the spin projections $|\pm 1\rangle$ are utilized for non-local gates instead of projections $|0,+1\rangle$ or $|0,-1\rangle$ [20], which can in principle increase the distance by a factor of 1.6.

The established method to produce NVs with spatial precision on a scale of tens of nm is the implantation of nitrogen ions into diamond and subsequent conversion to NVs in an annealing process [23]. In fact, all previously reported pairs of NV defects exhibiting dipolar coupling [20-22] have been created using this process in different variations. Yet, despite advances in the precision only two strongly coupled pairs could be identified so far [21,22], showing the probabilistic nature and low success rate of the process.

The technique is based on accelerated nitrogen ions directed towards a diamond sample [24] with a kinetic energy $E$, as shown in Figure 1 (b). Upon impact on the surface they penetrate the lattice and scatter on the carbon atoms, creating vacancies along their tracks, until all kinetic energy is consumed and they come to a stop either as substitutional or interstitial defects. During the annealing step the former can be converted to NV defects by capturing mobile vacancies that typically originate from lattice damage in the ion tracks. The resolution, or spread of stopping positions, is influenced by the accuracy of the ion beam, i.e. the distribution of surface penetration points. It can be reduced by focusing techniques [25] or a variety of mask techniques where nanometer sized channels allow ions to pass a medium in defined locations while otherwise being blocked, e.g. nano-channels in mica [26], e-beam lithography on polymers [27,28], or pierced AFM tips [29]. While ions can be implanted through a single channel (Figure 1 (b) left) leaving exact impact positions to chance, the ideal approach to engineer structures of coupled NV defects is to position ions individually through multiple channels (Figure 1 (b) right). However, even with a perfect point source on the diamond surface the accuracy is fundamentally limited by the ion scattering in the lattice (straggling). Higher ion energies will result in larger straggling ranges, as shown in Figure 1 (c). For example ions with energies higher than about 10 keV have straggling beyond 20 nm (FWHM), resulting in median couplings smaller than a few kHz (Figure 1 (d)). Consequently, straggling favors low ion energies for the creation of dipolar coupled NV defects.

The annealing process serves to convert implanted nitrogen atoms to NV defects and heal lattice damages induced by the implantation. However, as vacancies can also bind together to form divacancy defects or larger agglomerates, not all damages are repaired which can introduce decoherence on proximal NV spins, impair the charge state or prevent the formation of NV defects in the first place. Ideally an ion should stop far away from most of the lattice damage and vacancies should have a low density in order to avoid their agglomeration. As this is only the case at high energies (~MeV), coherence times of implanted NV decrease rapidly for lower energies. For example, NV created by 5 keV implantation have coherence times on the order of a few μs [30] while NVs created in different processes at the same depth, for example by delta-doping, can in principle reach hundreds of μs [31]. The conversion yield of ions to NV defects also decreases for lower energies [32], as shown in Figure 2 (a). Hence, to create two NVs at close distances, either a high ion fluence has to be used, causing more implantation damage and shorter coherence times of the resulting qubits, or many sites have to be implanted with few ions resulting in a low probability to create two nearby NV defects. In addition, lower ion energies result in a shallow implantation, where proximity to the surface exposes NV defects to noisy environments of electron spins deposited in layers on the diamond [33,34]. Thus, the quality of qubits poses a challenge to the implantation coupled pairs at lower energies. This especially impairs the multi-channel approach which requires low ion energies to retain the accuracy.

The counteracting effects of yield, decoherence and straggling suggest that the creation of coupled NV defects using current processing methods has generally low success rates. At high energies, NV defects with long coherence times can be created with high probability, but straggling makes large dipolar coupling strengths unlikely. On the other hand, at low energies ions can be positioned reliably at close distances, but the coherence times of resulting NVs are likely insufficient to detect any couplings to their neighbors. The combination of these trade-offs suggests that the highest success rates to produce strongly coupled NVs are to be found in an intermediate energy range. In the following we quantify the interrelations in simulations for single-channel implantation, present experimental results of implanted coupled NV pairs and investigate the potential scaling through ion implantation.

## 2. Estimation of optimal implantation parameters for coupled defects

In order to estimate the optimal implantation parameters and the success rate of generating strongly coupled pairs of NV defects through a single channel, we conduct a Monte-Carlo study of the implantation process. The parameters of interest are the channel aperture (diameter $d$ and area $A = d^2 \cdot \pi/4$), the ion energy $E$ and the ion fluence $F = n_{N^+}/A$, denoting an amount of nitrogen ions $n_{N^+}$ over an area $A$.

The ideal channel is a point source. In order to simulate a realistic case we choose a finite size. The highest success rates are expected for energies higher than a few keV. In these cases, the straggling is already on the order of tens of nm and aperture sizes of similar dimensions do not significantly reduce the overall position accuracy (see Figure 1 (c)). Hence, a diameter $d = 50$ nm that is technically feasible, e. g. via e-beam lithography in PMMA [27,28], is used.

The study is based on the strong coupling condition. We seek to find the ratio of implantation sites where at least two NV defects are created that fulfill this requirement. Thereby we need to model the amount of created NV defects, their coupling strength and their coherence times. In a single simulation run, the amount of ions $n_{N^+}$ passing the channel is drawn from a Poissonian distribution with a mean value $\bar{n}_{N^+}$ depending on the fluence $F$ and the aperture area $A$:

$$\bar{n}_{N^+} = F \cdot A = F \cdot \frac{\pi}{4} d^2. \tag{5}$$

Each ion receives a random lateral starting coordinate $\vec{r}_{\text{mask}}$ on the area of the aperture. Under the assumption of a homogeneous fluence throughout the aperture the whole area has a uniform probability of being hit by an ion. Afterwards the scattering of the ion in the diamond is simulated using stopping and range of ions in matter [35] where each ion receives a random displacement coordinate $\vec{r}_{\text{lattice}}$ based on its kinetic energy $E$. The stopping position is then obtained from the sum

$$\vec{r} = \vec{r}_{\text{mask}} + \vec{r}_{\text{lattice}}. \tag{6}$$

The amount of nitrogen ions that convert to NV defects during the annealing process can be modeled based on the empirical conversion yield data presented by Pezzagna, et al. [32] and shown in Figure 2 (a). The yield is assumed to be independent of the density of implanted ions, i.e. the ion fluence and each implanted ion has the same probability $y(E)$ to be converted into an NV defect that only depends on the energy $E$. The axis orientation for each NV defect is chosen at random from the set of four available crystal axes

$$\vec{a}_{\text{NV}} = \frac{1}{\sqrt{3}} \left\{ \begin{pmatrix} 1 \\ 1 \\ 1 \end{pmatrix}, \begin{pmatrix} 1 \\ -1 \\ -1 \end{pmatrix}, \begin{pmatrix} -1 \\ 1 \\ -1 \end{pmatrix}, \begin{pmatrix} -1 \\ -1 \\ 1 \end{pmatrix} \right\} \tag{7}$$

with equal probability. At this point all information to evaluate the dipolar coupling strength $\nu_{\text{dip}}$ between pairs of NV defects is available. The angular terms are obtained by the dot products

$$\cos \phi_{AB} = \vec{a}_A \cdot \vec{a}_B, \tag{8}$$

$$\cos\phi_{ir} = \frac{\vec{a}_i \cdot (\vec{r}_A - \vec{r}_B)}{r} \quad |_{i=A,B} \tag{9}$$

and the distance $r = |\vec{r}_A - \vec{r}_B|$.

The coherence times $T_2$ of implanted NV defects depend on many dynamics, e.g. spin-noise introduced by paramagnetic vacancy clusters from the implantation, substitutional and interstitial nitrogen defects, $^{13}C$ in the diamond lattice, surface spins and electric noise caused by unstable charge environments. As there is no complete theoretical framework that can reliably predict the coherence times of implanted NV defects we have to rely on empirical data to estimate the outcome depending on the ion energy $E$ and the fluence $F$. The used data was collected throughout the characterization of several samples implanted both through PMMA nano-apertures and without mask, and recorded as NV ensembles, at different energies and fluences. An observational model of the median coherence times using a logistic growth over the ion energy and an exponential decay over the ion fluence is fitted to the data and is shown in Figure 2 (b). Further details can be found in the supplementary material. For each NV defect taken in the Monte Carlo simulation a coherence time is drawn from an exponential distribution with the appropriate median value. Finally, the strong coupling condition of defect pairs can be evaluated.

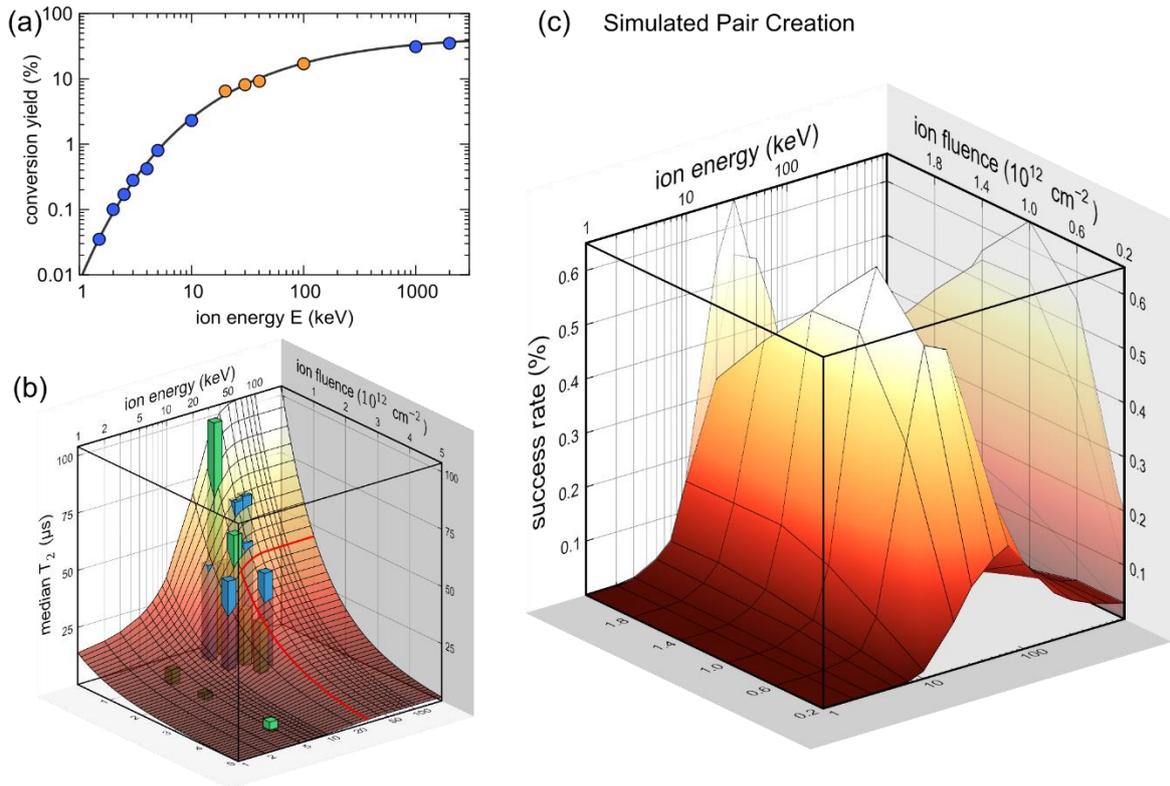

**Figure 2 (a)** Conversion yield of implanted nitrogen ions to NV defects depending on ion energy. Data taken in parts from S. Pezzagna, et al. [32] (blue) and complemented (orange). **(b)** Median coherence times $T_2$ of implanted NV spins depending on ion energy and ion fluence. The model is based on empirical data from single nanomask NVs (green bars) and NV ensembles (blue bars). The optimal simulated parameters in subfigure (c) have a median coherence times of 41 μs (red guides). **(c)** Simulated success rate of implanting pairs of NV defects through a 50 nm aperture where the dipolar coupling strength is larger than the decoherence rates of both spins. A maximum probability of 0.6% is reached at an energy of 30 keV and a fluence of $10^{12}$ cm$^{-2}$ corresponding to an average of 20 implanted ions within one site.

The results for several ion energies and fluences, each sampled over one million runs, is shown in Figure 2 (c). As expected, the highest success rates of implanting strongly coupled pairs of NV defects are found in an intermediate range of 20-50 keV. The maximum efficiency of 0.6 % is found at $E = 30$ keV and $F = 10^{12}$ cm$^{-2}$, corresponding to around 20 ions per site with an aperture area of $\pi/4 \cdot 50^2$ nm² and an average of 1.6 created NV per site with an expected yield of 8 % for an energy of 30 keV. Lower fluences will reduce the probability for two NV defects to be created and higher fluences result in shorter coherence times. The energy range below 5 keV that is suitable for a multi-channel approach, has success rates on the order of 0.025 %. Even when the overall coupling strengths are increased by simulating a point shaped aperture the highest success rates at 5 keV merely reach values of 0.09 %.

### 3. Experimental realization of coupled defects

With simulated probabilities below 1% the process is far from deterministic. Post-selecting sites from a set of several hundreds or thousands sites is, however, realistic. We therefore apply the simulated implantation parameters and prepare two identical [100]-oriented, single-crystal, CVD-grown diamond samples with natural abundance of $^{13}$C (1.1%). To form the implantation mask, we employed e-beam lithography technique using PMMA positive resist. An acceleration voltage of 30 kV and an aperture size of 10 μm were used to expose approximately 5,000 sites per sample with a diameter of 50 nm in the spin-coated PMMA (50K 9%) with the thickness of approximately 400 nm. Subsequently, $^{14}$N$^+$ ions are implanted with the calculated fluence of $10^{12}$ cm$^{-2}$ and energy of 30 keV with 0° incident angle (see Figure 3 (a)). After removing the PMMA masks the samples are annealed for two hours at 950°C in a $10^{-7}$ mbar vacuum. The carbon lattice is terminated with oxygen by boiling the diamond samples in acid in order to have the implanted NV defects in the negative charge state.

Figure 3 (b) shows an exemplary confocal fluorescence scan of NV sites implanted through the grid of nano-apertures. The amount of created NV defects per implantation site is assessed by their fluorescence intensity. We find a Poissonian distribution with an average of 1.6 NV defects per site (1.69 on sample A, 1.54 on sample B, see Figure 3 (c)), proving that the experimental parameters match the simulation. Throughout the characterization no significant difference between the two samples was found.

Each implantation site that is bright enough to host two or more NV defects is investigated for dipolar coupling. The spin resonances of parallel oriented NV spins overlap and thus individual addressing in double electron-electron resonance (DEER) experiments is challenging. Hence, one quarter of potentially coupling pairs are not identified. The coherence times for each addressable NV are determined by conducting electron spin echo envelope modulation (ESEEM) measurements in an axially aligned magnetic field of 5 mT. A histogram of measured coherence times is shown in Figure 3 (d), where an exponential distribution with a median $T_2$ of 33 μs (41 μs expected) can be seen. If a measured coherence time surpasses a threshold of 20 μs, corresponding to a decoherence rate of 50 kHz, the NVs are tested for dipolar coupling in a DEER experiment. For a detailed description of ESEEM and DEER experiments see the supplementary information.

For the characterization of a large amount of implantation sites an automated routine is employed. This has the advantage that implantation sites can be consistently investigated within an average measurement time of five minutes and without subjective human bias. On the other hand we expect the routine to produce false negatives without any indication of the failure rate. For example, the evolution times in DEER experiments were only measured for up to 100 μs, setting a lower bound of around 10 kHz for the detectable coupling. Positive results are always subject to validation in separate experiments.

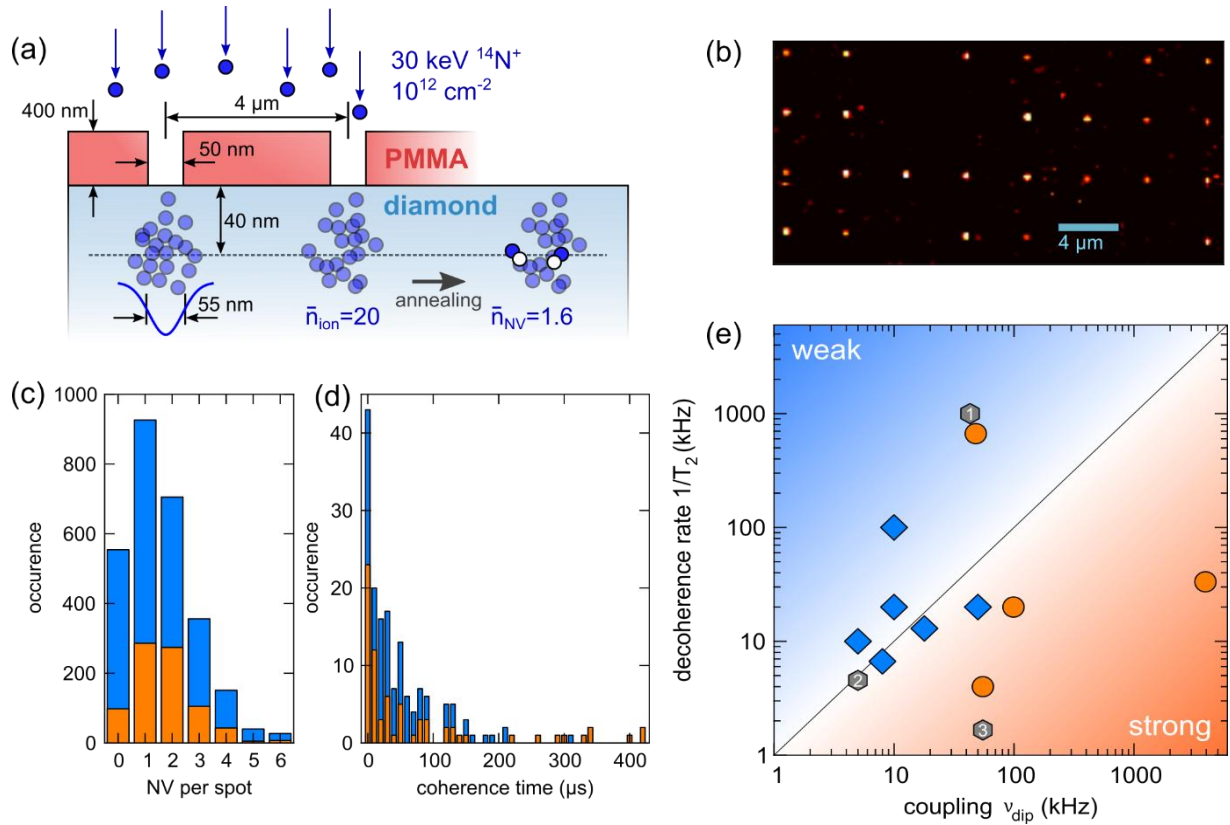

**Figure 3 (a)** Schematics of the implantation of $^{14}N^+$ ions through 50 nm apertures in a 400 nm PMMA mask. The spacing between holes is 4 μm. The simulated optimal parameters are used. This results in 20 implanted ions and two created NV defects on average, a depth of 40 nm and a resolution of 55 nm (FWHM). A total of 10,000 sites were implanted throughout two samples. **(b)** Confocal microscopy image of an implanted region with a grid of NV sites. **(c)** Histogram of amount of implanted NV defects per site counted in 818 and 1941 sites on sample A (orange) and B (blue), respectively. The average amount is 1.58 NVs per site. **(d)** Histogram of $T_2$ coherence times (ESEEM) of implanted NV recorded on 77 and 96 sites on sample A (orange) and B (blue), respectively. The median coherence time is 33 μs. **(e)** Characteristics of identified coupled NV pairs. For each found pair, the limiting decoherence rate is plotted over the dipolar coupling strength. The diagonal line marks the threshold of the strong coupling condition. Weakly and strongly coupled pairs appear in the top-left and bottom-right regions, respectively. Pairs found in sample A are marked as orange circles, sample B as blue diamonds. Previously reported pairs from P. Neumann, et al. (1), F. Dolde, et al. (2) and T. Yamamoto, et al. (3) are marked as grey hexagons.

The search was conducted over 6,000 implantation sites, amassing a total measurement time of about three weeks. Throughout both samples a total of six strongly coupled and four weakly coupled pairs were identified. The coupling strengths and limiting decoherence rates are shown in Figure 3 (e) and individual measurements are presented in the supplementary information. Most identified coupling strengths appear in the range of few tens of kHz. Lower couplings are hardly detected by the search routine. Only two couplings with 5 and 8 kHz could be identified in this range. The strongest recorded coupling strength is 3.9 MHz corresponding to maximum distance of 3 nm for an ideal angular constellation. The overall success rate of 0.1 % is within the same order of magnitude as the simulated 0.6 %. The actual amount of created pairs is expected to be larger because of omitted parallel NVs and false negatives from the search routine. No trimers or larger networks of coupled NV spins could be identified.

The presented experimental single-hole implantation has room for improvement. For example, the aperture size of 50 nm can be shrunk to smaller dimensions, decreasing the size of the positioning distribution from 55 nm (FWHM) by up to 45 % to 30 nm for a 10 nm hole. According to simulations this would increase the chance to create strongly coupled NV pairs by a factor of about 1.8. The coherence times of implanted NVs can also be improved, for example by using isotopically enriched $^{12}$C samples. However, hyperfine coupled $^{13}$C nuclear spin qubits might be an important resource for the scaling of the NV quantum register. Furthermore, it is doubtful that the $^{13}$C spin bath is limiting the NV coherence times for the desired low energy implantation range. The cause here are rather implantation damages and surface spins. Surface effects can be mitigated by growing an additional CVD layer after the implantation[30].

### 4. Further scaling

Although the statistic sample sizes are relatively small, the experiments show comparable results to the simulation. With this affirmation we investigate the probabilities for larger coupled networks using the same method. We define a network of size $n$ as any amount of NVs where $n$ defects are interconnected by strong coupling without interruption. More details can be found in the supplementary information.

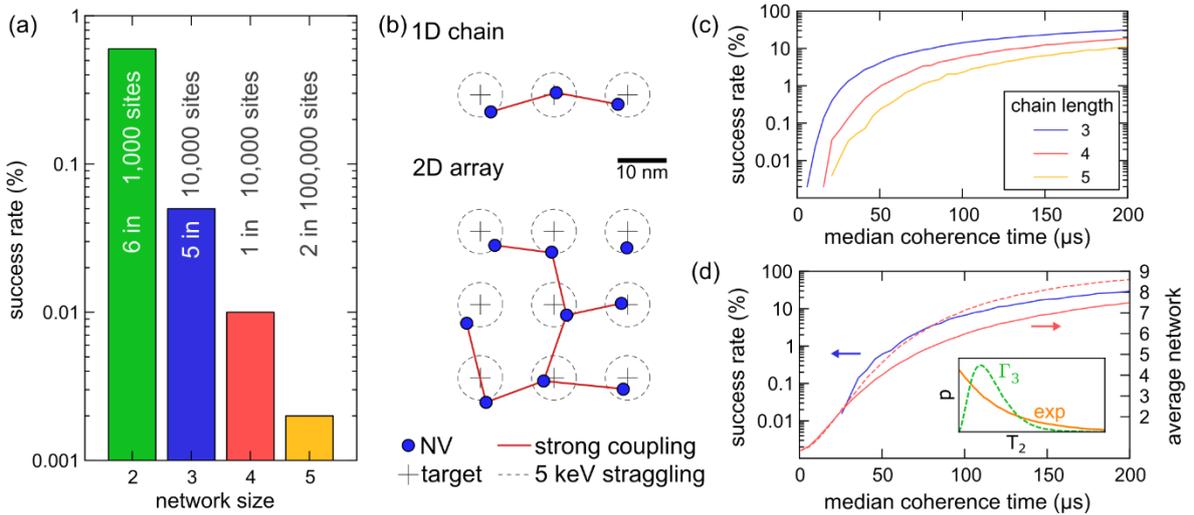

**Figure 4 (a)** Simulated success rates to implant strongly coupled networks with 2-5 nodes through a single 50 nm wide nano-channel. The same simulation method as in Figure 2 is used. **(b)** Exemplary structures of NV defects implanted at 5 keV through individual nano-apertures. The aperture spacing is 15 nm and the positioning is only limited by straggling (dashed lines). Assuming that exactly one NV is created for each channel, the strong coupling condition (solid red lines) is evaluated between assumed next neighbors. In this array example only eight of nine nodes are connected to a strongly coupled network. **(c)** Simulated success rates to implant strongly coupled NV spin chains of lengths 3-5 depending on the median of an exponential distribution from which coherence times are drawn. In this approach coherence times of tens to 100 μs are required for realistic success rates. **(d)** Simulated success rates (blue, left y-axis) to implant a strongly coupled 3x3 NV spin array depending on the same coherence time distributions and average resulting network size (red, right y-axis). Coherence times are drawn from exponential distributions (solid) and gamma distributions with shape-factor 3. The inset shows exemplary distributions.

The success rates to create triples, quadruples and quintuples are shown in Figure 4 (a), each for their respective optimal implantation parameters. For each additional node the probability decreases by about one order of magnitude. Extrapolating from the experiment a triple might require the investigation of an order of 10,000 implantation sites and months of measurement time to be found, whereas searching for any larger network appears unfeasible. One would expect that as the density of NV defects increases, the average distance decreases, favoring the creation of larger networks. However, in order to create the necessary densities fluences have to be increased and the decoherence rates outscale the coupling. In order to achieve realistic probabilities to scale the NV quantum registers either the coupling strengths at high energies or coherence times at low energies have to be increased. Straggling is inherent to implantation and unlikely to be circumventable, whereas coherence times of NV defects implanted at low energies may be improved through optimized processing methods during implantation and annealing[30,36,37].

Revisiting the multi-channel approach we conduct a theoretical study to estimate the necessary improvement in coherence times to make engineered NV spin chains and arrays feasible. In the ideal case exactly one NV is created for each nano-channel and the accuracy only depends on the fundamental straggling. In order to reach adequate coupling strengths of few kHz to tens of kHz, the distance between NVs should be within 20 nm. For this example an implantation site spacing of 15 nm and an ion energy of 5 keV used, which is technically feasible, for example with pierced AFM tips, or within reach with lithographically written nano-apertures. An exemplary one dimensional chain and a two-dimensional array of NV defects are shown in Figure 4 (b). In a Monte-Carlo simulation NV positions and axis orientations are distributed, coherence times are drawn from exponential distributions and the couplings between next neighbor NV nodes are calculated.

The spin chains are investigated for broken links. Figure 4 (c) shows the success rate depending on the median coherence time for different lengths. For this example coherence times have to reach 50 to 100 µs in order to achieve single percentage success rates. Coherence times of this magnitude are within reasonable reach for the chosen implantation energy [30], yet have still to be realized for fluences that could reliably produce an NV defect in each implantation site. A spin chain has also the drawback that each link has to be strongly coupled, which is reflected in the decreasing success rates of longer chains. In two dimensions each node can have several links and unconnected or broken nodes can still be tolerated.

The array is therefore investigated for the largest network of strongly coupled NV defects and compared to the success rate to achieve a network with all nine nodes connected, as shown in Figure 4 (d). While the success rate of the array scales similarly to the chains, requiring again at least 50 to 100 µs in order achieve single digit percentages, the average largest network size already reaches values of three to six nodes in the same range. Here the main reason for broken nodes or links are short coherence times that frequently occur. Assuming that with increased median values particularly short values can be avoided at the same time, we also simulate gamma distributions with a shape-factor of 3 ($\Gamma_3$) that have a gradual onset. At low median values there is no difference between the distributions. Around 30-50 µs, however, the $\Gamma_3$-distribution gains a significant advantage, which is the range where decoherence rates start scaling below the couplings. Hence not only overall coherence times of implanted NV defects have to be increased to a range that is already achieved with different generation approaches [30,31], but in particular short coherence times have to be avoided. With such improvements the probabilistic scaling of the NV quantum register appears to be feasible.

## 5. Conclusions

In conclusion we have devised a simulation framework that estimates the outcome of the implantation of coupled NV defects. Based on this we experimentally realized six strongly coupled pairs of NV defects, with coupling strengths of up to 4 MHz. While the success rate is far from ideal, our method signifies a substantial increase compared to previously reported techniques. Our recipe can be reproduced and may facilitate further research into dynamics of coupled solid-state spins, for example spatially correlated spectroscopy, entanglement protocols and quantum information processing in general.

While the presented Monte-Carlo simulations produced satisfactory results, they still rely on rudimentary models based on empirical data. The scattering of nitrogen ions in diamond is well understood. However, the dynamics of vacancies during the annealing process and the resulting conversion yield and coherence times of implanted NV defects lack comprehensive models. Progress in this field has the potential to improve the prediction results. Especially for future engineering of complex spin-chains and arrays, precise simulations may become invaluable tools to sieve through a vast parameter range in search for optimal implantation conditions.

Our work also shows that the creation process with contemporary methods is mostly probabilistic and generally showing success rates below 1% for pairs. While the precise positioning of NV defects certainly is important, especially for tailored constellations of coupled spins, the limiting factor for the scaling are the coherence times of implanted NV defects. Here processing methods need to increase the conversion yield and coherence times, for example through epitaxial overgrowth, in order to harness the benefits of already established precise ion positioning methods at low implantation energies.

**Acknowledgments**
We acknowledge financial support by the DFG (SFB-TR 21, SPP1601 and FOR1493), the EU (SIQS), the DIADEMS consortium and the MPG. Furthermore, we thank Wolfgang Brückner and Olga Lik for their support with the implantation.

# Supplementary information for
# "Efficient creation of dipolar coupled nitrogen-vacancy spin qubits in diamond"

I Jakobi, S A Momenzadeh, F Fávaro de Oliveira, J Michl, F Ziem, M Schreck, P Neumann, A Denisenko and J Wrachtrup

Email: i.jakobi@physik.uni-stuttgart.de

**Abstract**. In the following we give further detailed information on the used methods. We present the nitrogen-vacancy defect (NV) spin dynamics and dipolar coupling and the applied measurement protocols. We present additional detail of the characterized NV pairs. Furthermore we discuss the choices of statistical distributions and criteria for our simulations.

## 6. Single NV Hamiltonian

The Hamiltonian describing the relevant dynamics of a single negatively charged NV electron spin in its electronic ground state of a zero-field splitting term, a Zeeman-interaction term and hyperfine coupling terms to proximal nuclear spins:

$$\hat{H} = h\left( D\hat{S}_z^2 + \gamma \vec{B}\vec{S} + \sum_i \vec{S} A_i \vec{I}_i \right). \tag{1}$$

Here, $h$ is Planck's constant, $D = 2.87$ GHz is the zero-field splitting imposed by the crystal confinement of the electron spin, $\gamma = 28.03$ GHz/T is the gyromagnetic divided by $2\pi$ of the electron spin, $\vec{B}$ is the magnetic field in the location of the NV defect, $A_i$ are hyperfine tensors representing the coupling to different nuclear spins and $\vec{S} = (\hat{S}_x, \hat{S}_y, \hat{S}_z)^T$ and $\vec{I}_i = (\hat{I}_{x\,i}, \hat{I}_{y\,i}, \hat{I}_{z\,i})^T$ are the spin operators for the electron and nuclear spins respectively.

In the experimental conditions throughout our work the zero-field splitting is the dominating term and therefore defining the quantization axis $z$ along the NV symmetry axis, whereas all other terms can be regarded as perturbations in this basis.

External magnetic fields $\vec{B}$ couple through the Zeeman interaction to the spin. Here static fields mostly induce a Zeeman splitting along the quantization axis $z$, while oscillating fields can drive the spin between its eigenstates, giving rise to magnetic resonance experiments. The hyperfine coupling term accounts for the interaction with nuclear spins in the vicinity, for example the intrinsic nitrogen spin ($^{14}$N $I = 1$, $^{15}$N $I = 1/2$) or paramagnetic carbon in the diamond lattice ($^{13}$C $I = 1/2$). For one the interaction allows to control nuclear spins via the electron spin and is therefore essential to scaling of the single NV quantum register. In return the coupling to a bath of nuclear spins also gives rise to decoherence of the electron spin as it constitutes a channel through which quantum information stored on the electron spin state can be lost to the environment.

## 7. Dipolar interaction Hamiltonian

In the case of two NV electron spins the Hamiltonian is expanded by a second single spin Hamiltonian and a dipolar interaction Hamiltonian:

$$\widehat{H}_{\text{dip}} = \frac{\mu_0 \, h^2 \gamma^2}{4 \, \pi \, r^3} [\vec{S}_A \cdot \vec{S}_B - 3(\vec{S}_A \cdot \vec{r})(\vec{S}_B \cdot \vec{r})], \tag{2}$$

with $\mu_0$ being the vacuum permeability, $r$ being the distance of the spins and $\vec{r}$ being the unit vector linking them. Here the spin operators $\vec{S}'_i = (\hat{S}_{xi}, \hat{S}_{yi}, \hat{S}_{zi})^T \,|_{i=A,B}$, are defined in their proper coordinate frame $(x_i, y_i, z_i)$, given by the respective quantization axis, thus the operators have to be rotated into a common frame $(u, v, w)$ with rotation matrices $R_i$

$$\vec{S}_i = R_i \cdot \vec{S}'_i = \begin{pmatrix} R_{xu} & R_{yu} & R_{zu} \\ R_{xv} & R_{yv} & R_{zv} \\ R_{xw} & R_{yw} & R_{zw} \end{pmatrix}_i \begin{pmatrix} \hat{S}_x \\ \hat{S}_y \\ \hat{S}_z \end{pmatrix}_i \,|_{i=A,B} \,. \tag{3}$$

The spin operator expressions in the Hamiltonian can be rewritten as

$$\vec{S}_A \cdot \vec{S}_B = (R_A \cdot \vec{S}'_A) \cdot (R_B \cdot \vec{S}'_B), \tag{4}$$

$$(\vec{S}_A \cdot \vec{r})(\vec{S}_B \cdot \vec{r}) = \left((R_A \cdot \vec{S}'_A) \cdot \vec{r}\right)\left((R_B \cdot \vec{S}'_B) \cdot \vec{r}\right). \tag{5}$$

In general it is expected to find all possible combinations of the term

$$a_{ij} \hat{S}_{iA} \hat{S}_{jB} \,|_{i,j=x,y,z} \tag{6}$$

in the expanded dot products. However, as the Zeeman term usually dominates all terms except $i = j = z$ can be neglected:

$$\hat{S}_{iA} \hat{S}_{jB} \ll \gamma B \quad \forall \, i,j = x, y \tag{7}$$

The dot product in eq. 4 then reduces to

$$\vec{S}_A \cdot \vec{S}_B \approx \hat{S}_{zA} \hat{S}_{zB} \, (R_{zu\,A} R_{zu\,B} + R_{zv\,A} R_{zv\,B} + R_{zw\,A} R_{zw\,B}). \tag{8}$$

The term in the bracket is the dot product of the $z$ component vectors of $R_A$ and $R_B$, which can be expressed as

$$\hat{S}_{zA} \hat{S}_{zB} \, (\vec{R}_{zA} \cdot \vec{R}_{zB}) = \hat{S}_{zA} \hat{S}_{zB} \cos \phi_{AB} \tag{9}$$

and represents the projection of the quantization axes onto each other with their angle $\phi_{AB}$. The dot products in eq. 5 reduces accordingly to

$$(\hat{S}_A \cdot \vec{r})(\hat{S}_B \cdot \vec{r}) \approx \hat{S}_{zA} \hat{S}_{zB} \cos \phi_{Ar} \cos \phi_{Br} \tag{10}$$

assuming the vector $\vec{r}$ is defied in the same $(u, v, w)$ frame (without loss of generality as it can always be transformed into the same frame). With these terms the approximated Hamiltonian can then be written as

$$\widehat{H}_{\text{dip}} \approx \frac{\mu_0 \, h^2 \gamma^2}{4 \, \pi \, r^3} (\cos \phi_{AB} - 3 \cos \phi_{Ar} \cos \phi_{Br}) \, \hat{S}_{zA} \hat{S}_{zB}, \tag{11}$$

which is equivalent to the form shown in the main paper. Supplementary Figure 1 shows a schematic of the geometry. Note that a magnetic field dependent induced magnetic moment for initial spin state $m_S = 0$ is neglected, which can be assumed for operations restricted to $m_S = \pm 1$ or external fields close to zero, yet large enough to fulfil eq. 7. The angular constellations resulting in strongest coupling are axially aligned, parallel spins. Here all angels $\phi_{AB} = \phi_{Ar} = \phi_{Br} = 0$ and the angular term reduces to a factor of 2 (parallel) or -2 (antiparallel). The median angular contribution when all four crystal axes and all directions of $\vec{r}$ have equal probability is $\tilde{v}_{\text{dip}} \approx 0.7 \frac{\mu_0 \, h^2 \gamma^2}{4 \, \pi \, r^3}$.

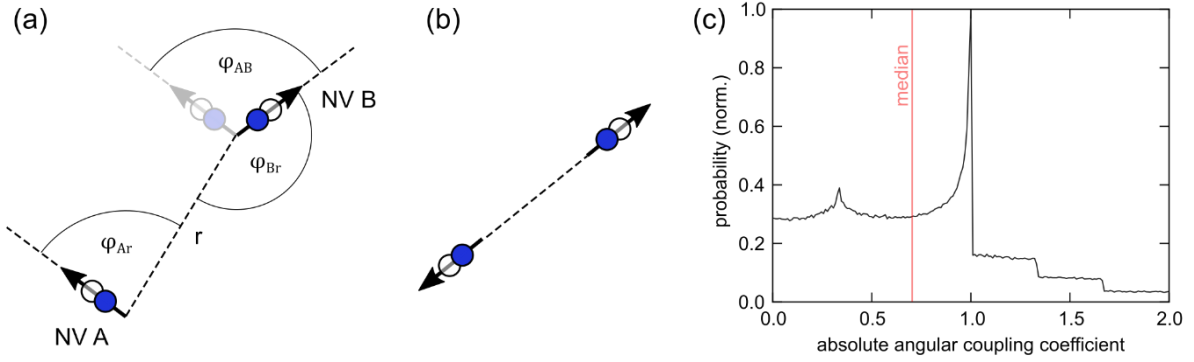

**Supplementary Figure 1 (a)** Schematics showing the relevant geometry for dipolar coupling of two NV electron spins. **(b)** The ideal angular constellations are axially aligned parallel or anti-parallel NVs. **(c)** Distribution of angular contributions to the dipolar coupling strength.

## 8. Coherence times measurements – Electron spin echo envelope modulation

The coherence times $T_2$ of NV electron spins are measured in Electron Spin Echo Envelope Modulation (ESEEM) measurements. Supplementary Figure 2 shows the corresponding measurement sequence. Here the electron spin is prepared in a superposition state $|\Psi\rangle = 1/\sqrt{2}\,(|0\rangle + e^{i\phi}|1\rangle)$ with a microwave $\pi/2$-pulse after initialization. In the subsequent evolution time $\tau$ the phase $\phi$ evolves with magnetic fields imposed by external sources, e.g. control fields and the spin bath, according to

$$\phi_1(\tau) = 2\pi\,\gamma \int_0^\tau dt\, B_z(t) \qquad (12)$$

The spin is flipped with a microwave $\pi$-pulse and left to evolve for another time $\tau$ where the phase follows

$$\phi_2(\tau) = -\,2\pi\,\gamma \int_\tau^{2\tau} dt\, B_z(t) \qquad (13)$$

Finally the total phase $\phi = \phi_1 + \phi_2$ is projected onto a population state with another $\pi/2$-pulse and read out. The intermediate $\pi$-pulse refocuses any phase evolution caused by magnetic field dynamics that do not change over the course of the measurement, effectively eliminating for example homogeneous broadening. What is left are phase evolutions caused by magnetic noise from the environment which will introduce a decay of coherence over the total evolution time $2\tau$, where the coherence time $T_2$ represents the characteristic decay length.

The interaction with the spin bath is in part coherent resulting in periodic decays and revivals of the ESEEM signal as can be seen in Supplementary Figure 2. The periodicity is caused by the Larmor precessions of the surrounding $^{13}$C nuclear spins and revivals occur every $\tau = \gamma_C \cdot |\vec{B}|$, where $\gamma_C = 10.7$ MHz/T is the gyromagnetic ratio divided by $2\pi$ of $^{13}$C nuclear spins. In the experiments magnetic fields of $|\vec{B}| = 5$ mT are used so revivals are expected to occur over the total evolution time every $2\tau = 37.4$ μs.

For the characterization of the implantation, the coherence times were only recorded on a subset of sites (sample A 77 NVs, sample B 96 NVs) with detailed ESEEM measurement. The search routine on the other hand merely measured the fluorescence contrast on the first and second revival to quickly estimate the decay.

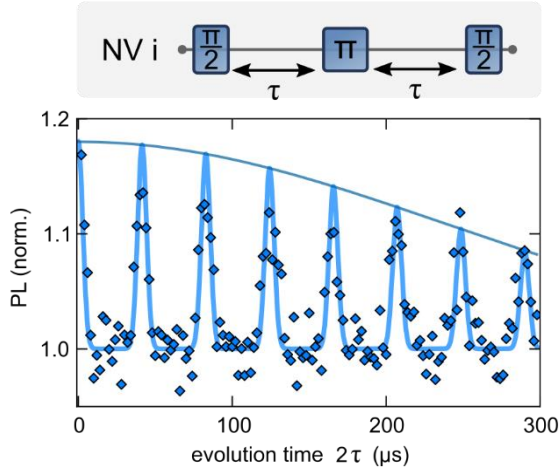

**Supplementary Figure 2** ESEEM pulse sequence and exemplary measurement.

In many cases the coherence times can be prolonged to a characteristic $T_2^\varrho > T_2$ by applying additional π-pulses through-out the evolution time that decouple the NV electron spin from higher orders of magnetic field dynamics from the environment. In this work we refrained from using these dynamical decoupling sequences. First they may not necessarily be compatible with actual quantum computing gate operations and second we seek to find the bounds of feasibility rather than achieve records based on circumstantial conditions. Hence we keep our study general as possible.

## 9. Dipolar coupling measurements – Double electron electron resonance

In order to detect dipolar coupling a measurement sequence as shown in Supplementary Figure 3 is applied. One NV takes the role of a sensor. After initialization it is controlled with microwave pulses to perform an ESEEM sequence with a magnetic field sensitive phase $\phi$ as described in the previous section. The second NV acts as emitter. It is prepared in an eigenstate and flipped by a microwave π-pulse during the evolution of the sensor. As the dipolar coupling strength $\nu_{\mathrm{dip}}(m_{S\,A}, m_{S\,B})$ between the two NV changes depending on the magnetic quantum numbers $m_{S\,i} = \hat{S}_{z\,i}|\Psi_i\rangle \;|_{i=A,B}$ , its contribution to the induced phase evolution of the sensor is not fully refocused. Hence the coupling strength $\nu_{dip}$ can be directly projected onto the phase evolution of the sensor by sweeping the point in time where the emitter is flipped.

In order to address both the sensor and emitter spin individually, selective microwave pulses have to be applied on the magnetic resonances. In a homogeneous magnetic field resonances of NVs with parallel axes overlap in general and can thus not be addressed individually. For this reason parallel NV are discarded from investigations.

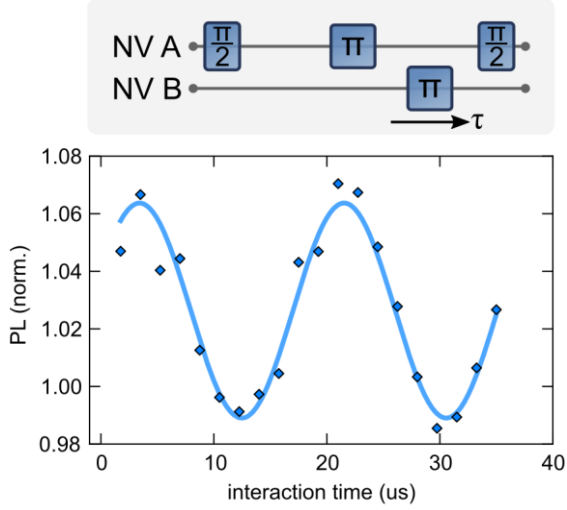

**Supplementary Figure 3** – DEER pulse sequence and exemplary measurement.

## 10. Empirical model of coherence times

In order to model the coherence times of implanted NVs we rely on empirical data. We record statistics of single NV coherence times on eight nanomask implantations and ensemble coherence times of eight implantations without mask. The median coherence times are shown in Supplementary Figure 4 (a). All diamond samples used are CVD-grown single crystals with a natural abundance of paramagnetic $^{13}$C. We apply a simple model to the data based on a few assumptions. At the highest energies and lowest fluences we expect coherence times to be limited by the $^{13}$C spin bath. Here we find the median coherence time $\tilde{T}_2$ to be on the order of 100 μs (mean $\bar{T}_2$ = 150 μs). For the lowest energies, surface effects will dominate the decoherence processes, where median coherence times typically are limited to hundreds of ns. In between we expect a logistic growth over the energy. For the fluence we assume an exponential decrease. While there indications that there is a saturation effect for lower fluences, as well, we have not enough data to support this. In addition we expect the pair creation probabilities in the regime of lower fluences to be limited by the amount of created NVs rather than their coherence times and thus minor discrepancies between variations of the model. The empirical model can then be expressed as

$$\tilde{T}_2(E,F) = c_0 \cdot \frac{1}{2}\left(1 + \tanh\frac{E - c_1}{c_2}\right) \cdot e^{-\frac{F}{c_3}}, \tag{14}$$

where the coefficient $c_i$ are determined in a fit to the data. We find a saturation value $c_0$ = 104 μs, a logistic inflection point $c_1$ = 14.8 keV, a curvature parameter $c_2$ = 16.3 keV and a decay constant $c_3$ = $1.2 \cdot 10^{12}$ cm$^{-2}$. The resulting median coherence times curve is show in Supplementary Figure 4 (b).

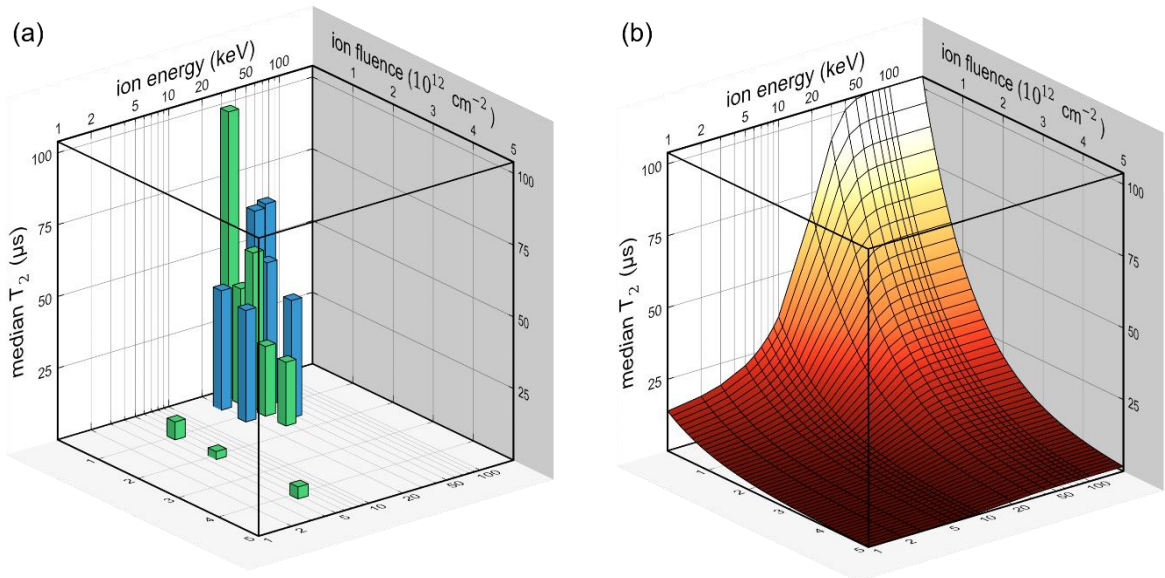

**Supplementary Figure 4 (a)** Empirical median coherence times of NVs implanted at different energies and fluences. The data was collected on and eight nanomask implantations (measured on single NV, green) and eight implantations without mask (measured on ensembles, blue). **(b)** Fitted median coherence times.

## 11. Coherence times distributions

For the simulations we typically use exponential distributions of coherence times. This shape of distribution is commonly found in NVs implanted with high fluences, for example in the data recorded for the coherence times model or as seen Figure 3 (d) in the main text. The shape of coherence times distributions may, however, be an important factor in the design of NV structures, as high probabilities for short coherences will impact the success rates of the strong coupling condition. For the simulation of the scaling we therefore also investigate gamma distributions of higher orders that would particularly reflect the avoidance of short coherence times. Gamma distributions with different shape factors $k$ are shown in Supplementary Figure 5. In the simulations we chose the shape of $k = 3$ as $k = 2$ distributions have steep onset at low values and therefore do not correspond to our requirements.

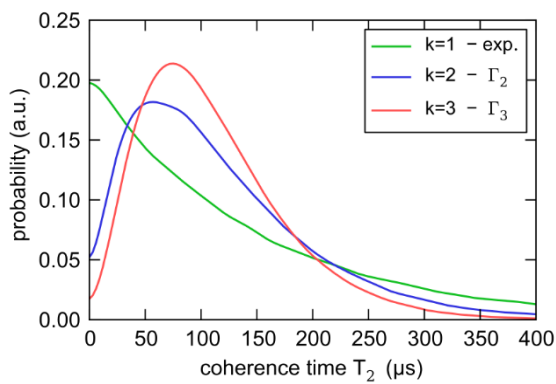

**Supplementary Figure 5** Comparison of statistical coherence time distributions for equal median coherence time of 100 μs. Exponential distributions (green) are typically observed in high fluence implantations and have a high probabilities for short coherence times. Gamma distributions of shape $k = 2$ (blue) and 3 (red) are suitable to model avoidance of short coherence times.

## 12. Detailed $T_2$ and $\nu_{\text{dip}}$ of identified coupled sites

The measurements of coherence times (ESEEM) and dipolar couplings (DEER) of identified sites with coupling NV spins are shown in Supplementary Figure 6 and a list of the derived values is shown in Supplementary Table 1.

**Supplementary Table 1** – List of sites with identified dipolar coupling and list NV characteristics.

| sample | site | $T_2$ of $NV_A$ (µs) | $T_2$ of $NV_B$ (µs) | Limiting $1/T_2$ (kHz) | $\nu_{\text{dip}}$ (kHz) | strong |
|---|---|---|---|---|---|---|
| **A** | 1 | 120 | 30 | 33 | 3900 | yes |
| **A** | 2 | 80 | 50 | 20 | 99 | yes |
| **A** | 3 | 400 | 250 | 4 | 55 | yes |
| **A** | 4 | 60 | 2 | 500 | 48 | no |
| **B** | 1 | 90 | 50 | 20 | 50 | yes |
| **B** | 2 | 120 | 80 | 13 | 18 | yes |
| **B** | 3 | 310 | 150 | 7 | 8 | yes |
| **B** | 4 | 190 | 50 | 20 | 11 | No |
| **B** | 5 | 90 | 2 | 500 | 10 | No |
| **B** | 6 | 120 | 80 | 13 | 5 | No |

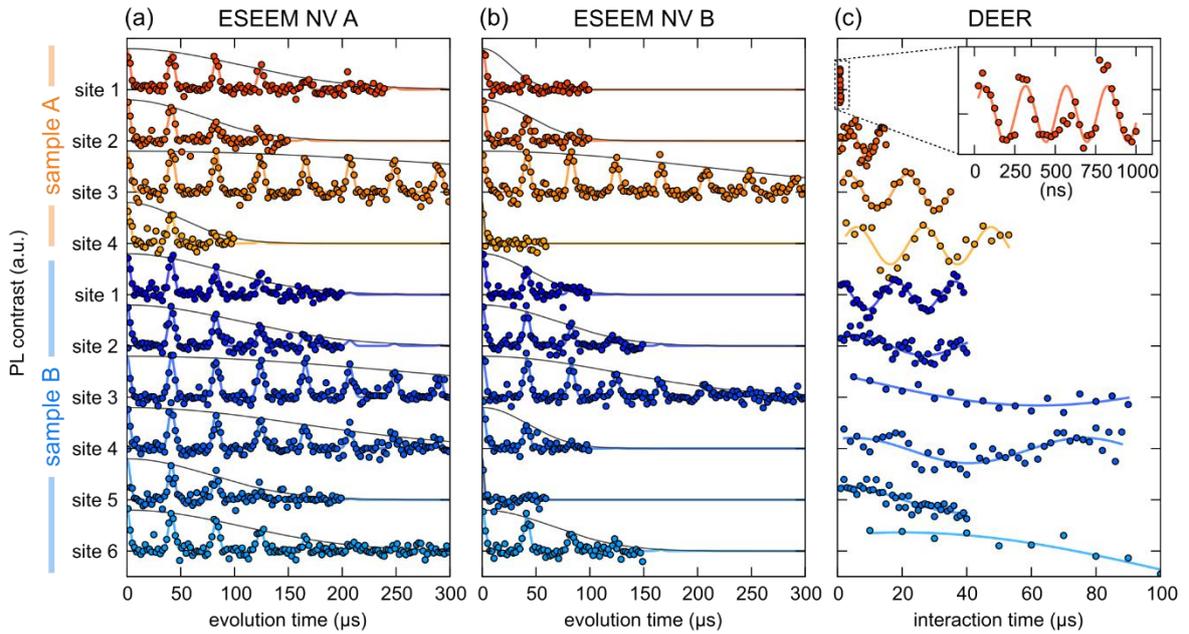

**Supplementary Figure 6** ESEEM of NV A **(a)**, NV B **(b)** with Gaussian envelope fits to derive $T_2$ and DEER **(c)** measurements with sinusoidal fits to derive $\nu_{\text{dip}}$ in sites with identified dipolar coupling.

## 13. Criteria for Coupled Networks

We consider a site with any amount of NV defects a coupled network size $n$ when a connected graph of the order $n$, where NV defects represent nodes and couplings fulfilling the strong coupling condition represent edges, can be found. This implies that the NVs do not have to be ordered in any fashion and that not every possible connection has to be strongly coupled. Furthermore a complete graph, where all possible pairs of nodes are connected, is equally considered a network as a linear graph or chain.

Supplementary Figure 7 shows all possible graphs for orders two through five that are considered a network of the respective size.

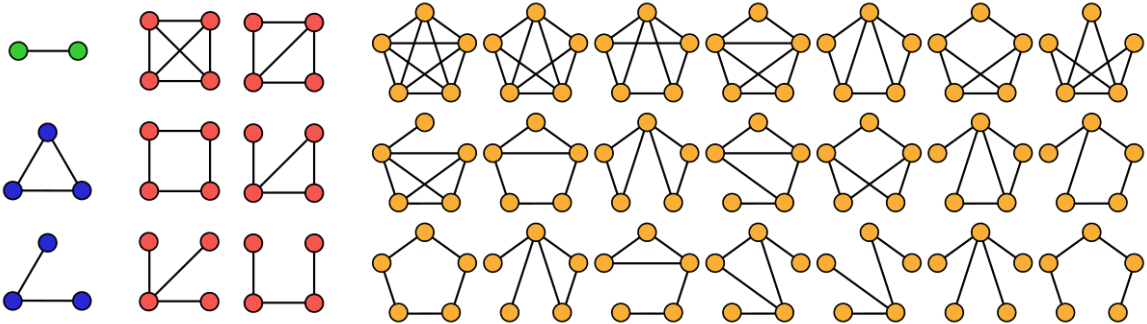

**Supplementary Figure 7** All possible graphs with two through five nodes that are considered a coupling network of the respective size. The nodes (circles) represent NV defects and the edges (lines) represent couplings between two NV that fulfil the strong coupling condition.